\pdfoutput=1

\documentclass[twocolumn,aps,prl,superscriptaddress,showpacs,floatfix]{revtex4-2}

\usepackage{graphicx,float}
\usepackage{epsfig}
\usepackage{epstopdf}

\usepackage{amsmath}
\usepackage{amssymb}
\usepackage{bm}
\usepackage{bbm}
\usepackage{dsfont}
\usepackage{bbold}
\usepackage{multirow}
\usepackage{color}
\usepackage[usenames,dvipsnames]{xcolor}
\usepackage[colorlinks=true,citecolor=blue,urlcolor=blue]{hyperref}

\usepackage{hyperref}
\hypersetup{colorlinks=true,linkcolor=blue,citecolor=blue,urlcolor=blue}

\newcommand{\be}{\begin{equation}}
\newcommand{\ee}{\end{equation}}

\newcommand{\sket}[1]{{\ensuremath{\lvert#1\rangle}}}
\newcommand{\lket}[1]{{\ensuremath{\left\lvert#1\right\rangle}}}
\newcommand{\ket}[1]{\if@display\lket{#1}\else\sket{#1}\fi}

\newcommand{\sbra}[1]{{\ensuremath{\langle#1\rvert}}}
\newcommand{\lbra}[1]{{\ensuremath{\left\langle#1\right\rvert}}}
\newcommand{\bra}[1]{\if@display\lbra{#1}\else\sbra{#1}\fi}

\newcommand{\sbraket}[2]{{\ensuremath{\langle#1\rvert#2\rangle}}}
\newcommand{\lbraket}[2]{{\ensuremath{\left\langle#1\!\left\rvert\vphantom{#1}#2\right.\!\right\rangle}}}
\newcommand{\braket}[2]{\if@display\lbraket{#1}{#2}\else\sbraket{#1}{#2}\fi}

\newcommand{\sketbra}[2]{{\ensuremath{\lvert #1\rangle\!\langle #2\rvert}}}
\newcommand{\lketbra}[2]{{\ensuremath{\left\lvert #1\right\rangle\!\!\left\langle #2\right\rvert}}}
\newcommand{\ketbra}[2]{\if@display\lketbra{#1}{#2}\else\sketbra{#1}{#2}\fi}

\begin{document}

\title{All-in-fiber dynamic orbital angular momentum mode sorting}

\author{Alvaro Alarcón}
\affiliation{Link\"{o}pings Universitet, Institutionen f\"{o}r Systemteknik, Link\"{o}pings Universitet, Link\"{o}ping 581 83, Sweden.}

\author{Santiago G\'{o}mez}
\affiliation{Universidad de Concepci\'{o}n, Departamento de F\'{i}sica, Universidad de Concepci\'{o}n, Casilla 160-C, Concepci\'{o}n, Chile.}
\affiliation{Universidad de Concepci\'{o}n, Millennium Institute for Research in Optics, Universidad de Concepci\'{o}n, Casilla 160-C, Concepci\'{o}n, Chile.}
\affiliation{Departamento de F\'{i}sica, Universidad del B\'{i}o-B\'{i}o, Collao 1202, Casilla 5C, Concepci\'{o}n, Chile.}

\author{Daniel Spegel-Lexne}
\affiliation{Link\"{o}pings Universitet, Institutionen f\"{o}r Systemteknik, Link\"{o}pings Universitet, Link\"{o}ping 581 83, Sweden.}

\author{Joakim Argillander}
\affiliation{Link\"{o}pings Universitet, Institutionen f\"{o}r Systemteknik, Link\"{o}pings Universitet, Link\"{o}ping 581 83, Sweden.}

\author{Jaime Cari\~{n}e}
\affiliation{Departamento de Ingenier\'{i}a El\'{e}ctrica, Universidad Cat\'{o}lica de la Sant\'{i}sima Concepci\'{o}n, Concepci\'{o}n, Chile.}

\author{Gustavo Ca\~{n}as}
\affiliation{Departamento de F\'{i}sica, Universidad del B\'{i}o-B\'{i}o, Collao 1202, Casilla 5C, Concepci\'{o}n, Chile.}

\author{Gustavo Lima}
\affiliation{Universidad de Concepci\'{o}n, Departamento de F\'{i}sica, Universidad de Concepci\'{o}n, Casilla 160-C, Concepci\'{o}n, Chile.}
\affiliation{Universidad de Concepci\'{o}n, Millennium Institute for Research in Optics, Universidad de Concepci\'{o}n, Casilla 160-C, Concepci\'{o}n, Chile.}

\author{Guilherme B. Xavier}
\affiliation{Link\"{o}pings Universitet, Institutionen f\"{o}r Systemteknik, Link\"{o}pings Universitet, Link\"{o}ping 581 83, Sweden.}
\email{guilherme.b.xavier@liu.se}

\begin{abstract}
The orbital angular momentum (OAM) spatial degree of freedom of light has been widely explored in many applications, including telecommunications, quantum information and light-based micro-manipulation. The ability to separate and distinguish between the different transverse spatial modes is called mode sorting or mode demultiplexing, and it is essential to recover the encoded information in such applications. An ideal $d$ mode sorter should be able to faithfully distinguish between the different $d$ spatial modes, with minimal losses, have $d$ outputs, and have fast response times. All previous mode sorters rely on bulk optical elements such as spatial light modulators, which cannot be quickly tuned and have additional losses if they are to be integrated with optical fiber systems. Here we propose and experimentally demonstrate, to the best of our knowledge, the first all-in-fiber method for OAM mode sorting with ultra-fast dynamic reconfigurability. Our scheme first decomposes the OAM mode in fiber-optical linearly polarized (LP) modes, and then interferometrically recombines them to determine the topological charge, thus correctly sorting the OAM mode. In addition, our setup can also be used to perform ultra-fast routing of the OAM modes. These results show a novel and fiber integrated form of optical spatial mode sorting that can be readily used for many new applications in classical and quantum information processing. 
\end{abstract}

\maketitle

\section{Introduction}
The orbital angular momentum modes (OAM) of light \cite{Allen:92, Shen:19} have captured the interest of the scientific community due to their many possible applications, such as optical manipulation \cite{Grier:2003, Padgett:11}, information transfer \cite{Gibson:04}, data multiplexing \cite{Wang:12}, quantum key distribution \cite{Mirhosseini:15} and high-dimensional quantum information processing \cite{Erhard:18}. OAM modes are commonly represented using a Laguerre-Gaussian (LG) basis set, which has a ring-like amplitude profile and azimuthal phase dependence $\exp(i\ell\phi)$. LG modes are characterized by $\ell$ azimuthal and $p$ radial integer numbers \cite{Allen:92}. In contrast, the modes that only carry orbital angular momentum (OAM) have only an azimuthal phase structure and are described only by the index $\ell$. The fact that OAM$_{\pm\ell}$ modes are mutually orthogonal and can generate an infinite state space makes them a suitable platform for applications requiring $d$-spatial modes (or $d$-dimensions), such as classical and quantum communications. For many applications it is crucial to develop OAM-supporting devices and systems that are compatible with optical fibers, can be integrated into other systems and are scalable. Direct compatibility with optical fibers was not an issue in the past, but since the advent of space-division multiplexing (SDM) optical fibers, which can support the propagation of OAM modes \cite{Xavier:20}, the use of bulk optical elements can be a major disadvantage.\\

\begin{figure}[htbp]
\centering
\includegraphics[width=0.45\textwidth]{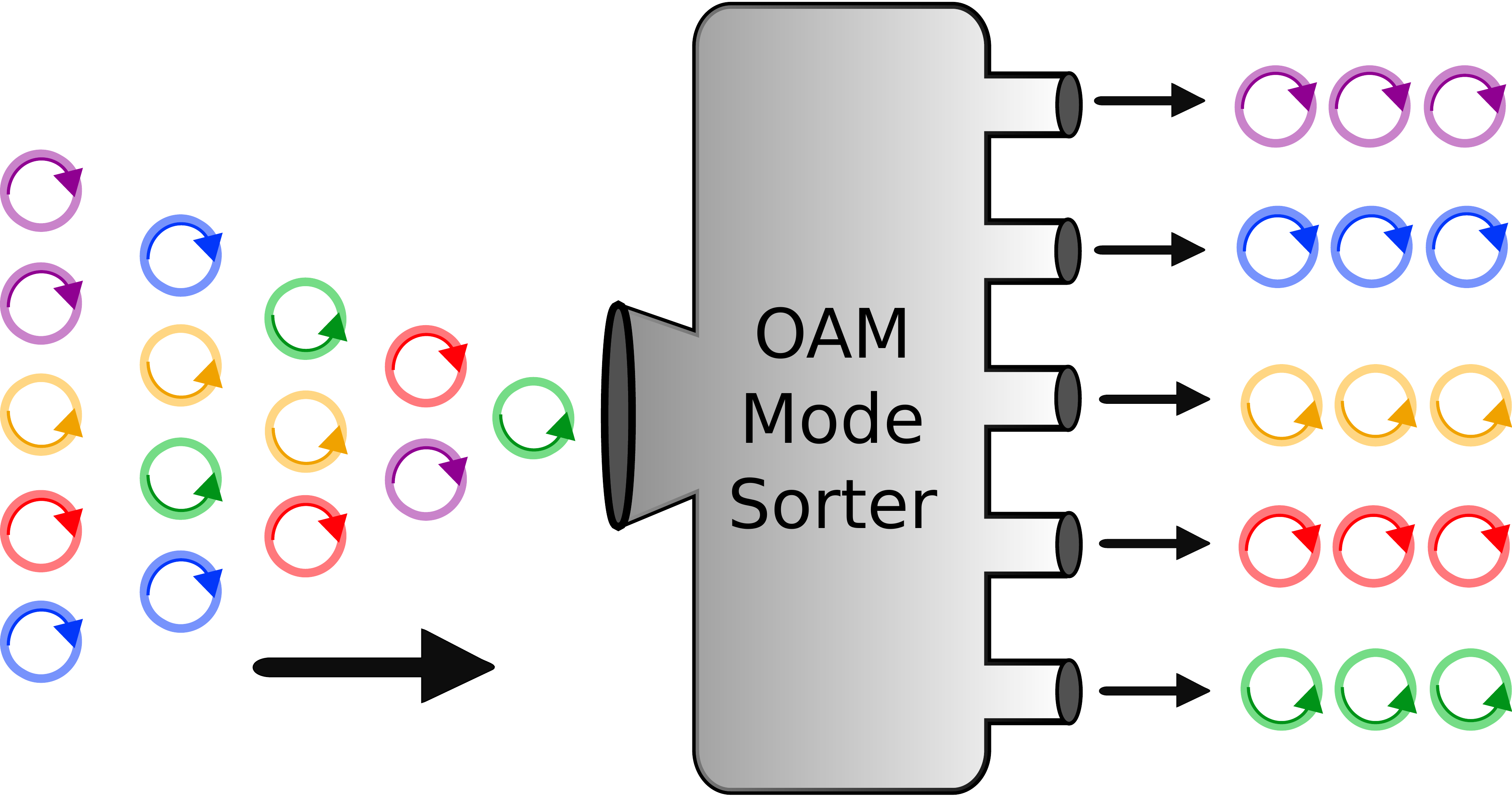}
\caption{General ideal OAM mode sorter. The purpose of an OAM mode sorter is to sort incoming OAM modes to a predetermined output, which depends on both the topological charge of the incoming beam and any operations carried out by the sorter.}
\label{Fig1}
\end{figure}

Although some efforts have been made to generate OAM modes using fiber systems \cite{Li:15, Chang:22, Wu:22, LIAN:22, Alarcon:23, Lu:23, Heng:18}, the efficient detection of OAM modes using all-in-fiber platforms remains an ongoing challenge. An OAM mode sorter is a device capable of multiplexing different incoming electromagnetic fields that carry OAM. Its internal operation depends on the spatial information carried by the incoming mode \cite{Fontaine:19, Ariyawansa:21}. Fig. \ref{Fig1} schematically shows the behavior of a general OAM mode sorter, where several OAM modes with different azimuthal indexes can be sorted at the output of the system.\\

OAM mode sorters have been implemented using several techniques, such as with the use of spatial light modulators \cite{Jia:22, Fontaine:19, Fickler:20, Lavery:12}, q-plates \cite{Jia:19, Karimi:09}, geometrical transformations \cite{Huang:15}, log-polar coordinate transformations \cite{Gregorius:10, Dudley:13}, spiral transformations \cite{Wen:18}, or multiplane converters \cite{Kupianskyi:23, Zhang:20, Morizur:10}. There are also free-space interferometric-based schemes that perform OAM mode sorting by applying relative phases related to the topological charge $\ell$ of the incoming modes using dove prisms \cite{Leach:02, JoL:2004, Cozzolino:19}. All these systems are based on performing operations on OAM modes using bulk optics, which makes them difficult to integrate into current telecommunications networks. Furthermore, when it comes to scaling up the number of spatial modes to be distinguished, this type of scheme becomes complex. A fiber-based technique recently presented is a sorter using optical fiber delays \cite{Bromberg:22}, taking advantage of the fact that the different modes propagating in a fiber will separate due to modal dispersion. This method has a limited bandwidth due to the long delays needed. Finally, a static multiplexer/demultiplexer based on a photonic lantern has been presented \cite{Eznaveh:18}, however as it has no reconfigurable capabilities, its applications in quantum information processing or in dynamic optical networks are impractical.\\

In this paper, we present a novel all-in-fiber platform, with ultra-fast reconfigurability, for interferometric sorting of arbitrary incoming OAM$_{\pm\ell}$ modes. Our approach is based on the observation that any OAM$_{\pm\ell}$ mode can be decomposed into a combination of an odd and an even linearly polarized (LP) mode, with a relative phase difference of $\pm  \pi/2$ \cite{Ma:12, Zeng:18}. The central components of our platform consist of an optical fiber that can support multiple transverse spatial propagation modes and a photonic lantern (PL) \cite{Leon-Saval:10, Birks:15}. We can immediately obtain complete information about the parameter $\pm\ell$ of the incoming OAM mode by detecting which output of our system the light beam exits through. Our scheme can be scaled up in dimensionality by employing components that support more spatial modes. In addition, we can operate our system from a passive mode, where the modes are sorted, to an active mode where we show that the OAM mode information can be arbitrarily redirected to another route. To experimentally demonstrate the feasibility of the scheme, we successfully implement an all-in-fiber reconfigurable mode sorter for the OAM$_{+1}$ and OAM$_{-1}$  modes. As we have used off-the-shelf components, our system can be easily integrated into fiber-optical telecommunication networks. Our sorter can operate at high speeds, being able to perform routing operations with a switching time as low as $7$ ns, limited by the driving electronics. To the best of our knowledge, this represents a record in speed response for OAM mode sorters, making it an excellent candidate for active (or passive) demultiplexing systems in current optical networks and for applications in quantum information processing.

\section{All-in-fiber OAM mode sorter}\label{sec:All-fiber OAM mode ssorter}

\begin{figure*}[ht]
\centering
\includegraphics[width=1.02\textwidth]{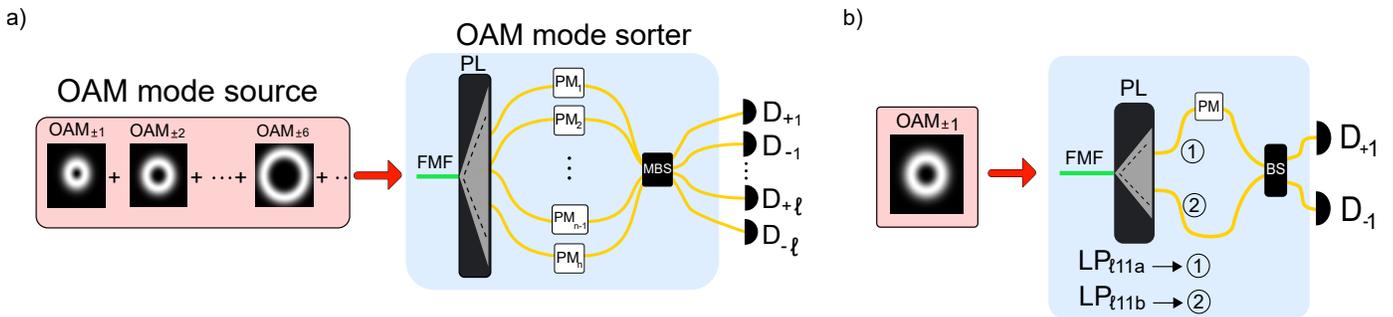}
\caption{The proposed OAM mode sorter. a) In this general high-dimensional case, there is a source emitting OAM modes with different azimuthal indices. Each OAM mode is mapped onto different output paths from the photonic lantern (PL), which are then superposed by a multiport beam splitter (MBS). Each output path is connected to a detector to determine the topological charge of each OAM mode. b) The experiment realized on the two-dimensional case allowing the sorting of the OAM$_{+1}$ OAM$_{-1}$ modes, with detectors placed at the two outputs. D: Detector; FMF: Few-mode fiber; PL: Photonic lantern; PM: Phase modulator.}
\label{Fig2}
\end{figure*}

In Figure \ref{Fig2}a we present the scheme of our proposal for sorting OAM states. The inputs to our scheme are any state between $\text{OAM}_{+\ell}$ or $\text{OAM}_{-\ell}$, where $\ell$ can only take integer values. The mode sorter consists of a few-mode fiber \cite{Sillard:14} connected to a photonic lantern \cite{Velazquez-Benitez2018}. 
The physical principle behind our experiment is the fact that any OAM$_{\pm \ell}$ mode within a few-mode fiber can be decomposed into the following LP modal superposition \cite{Alarcon:23, Zhang:19, Lu:18, Zeng:18}: 
 \begin{equation}
 \text{OAM}_{\pm \ell} = \text{LP}_{\ell 1 (\text{even})} \pm i \cdot \text{LP}_{\ell 1 (\text{odd})}
 \label{eq1}
\end{equation}

The photonic lantern maps the LP modes that follow the above decomposition from a specific OAM$_{\pm \ell}$ mode coming from the few-mode fiber with a one-to-one correspondence into a pair of single-mode fiber outputs only supporting the fundamental LP$_{01}$ mode (Gaussian). The lantern also preserves the phase relation between the LP mode pair. The single-mode fibers are then recombined interferometrically with a multi-port beam splitter (MBS) \cite{Carine:20}. Phase modulators (PMs) are placed in each arm of this high-dimensional interferometer to adjust the internal relative phases. It is therefore possible to identify uniquely the state's topological charge $\pm\ell$ by detection at the output of the interferometer. This method is completely in-fiber, and this makes it perfect for integration to OAM sources and systems which are already done fully in-fiber, which is becoming more common \cite{Xavier:20, HENG:17, LIAN:22, Alarcon:20}. Observe that if the appropriate phases are set in all the phase modulators \cite{Carine:20} it is possible to route the mode information to any output. Furthermore, this reconfigurable setup allows one to implement any unitary transformation onto a high-dimensional OAM encoded quantum state, which is highly sought after in quantum information. In this work we perform an experimental demonstration of the scheme to sort the OAM$_{+1}$ and OAM$_{-1}$ modes (Fig. \ref{Fig2}b), based on 3-mode few-mode fibers and photonic lanterns. 

\section{Experimental setup}  

\begin{figure*}[!t]
\centering
 \includegraphics[width=1.02\textwidth]{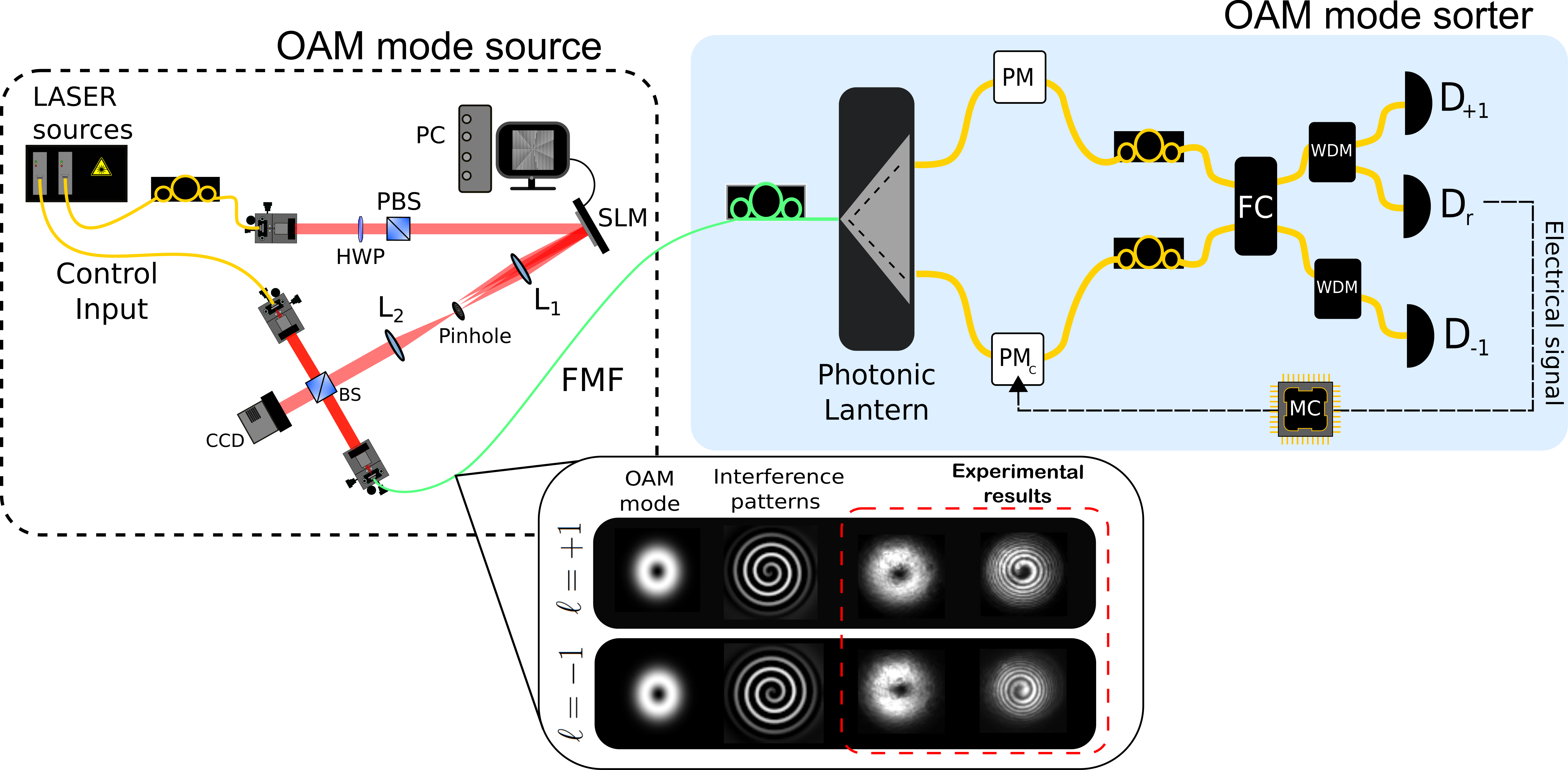}
 \caption{Experimental setup. The setup for demonstrating the sorting of OAM$_{\pm1}$ modes can be divided into two main parts: the OAM$_{\pm1}$ mode source and the OAM$_{\pm1}$ mode sorter. The OAM source can switch between OAM$_{+1}$ and OAM$_{-1}$ by changing the forked diffraction grating on the spatial light modulator (SLM). The mode sorter is an all-fiber interferometer based on space-division-multiplexing fiber technology. The computer-generated holograms used to create the OAM$_{\pm1}$ are a superposition of a helical phase and a diffraction grating. The inset shows the generated spatial intensity and phase profiles (interference patterns) of the OAM modes. The phase profiles are obtained from the interference pattern between the corresponding OAM mode and a Gaussian beam to distinguish the topological charge $\ell=\pm1$. These spatial patterns are detected by an InGaAs CCD camera. A control laser at a different wavelength is injected onto the FMF through the beamsplitter (BS) before the CCD camera and is used as feedback signal for a microcontroller (MC) running a control algorithm to stabilize the phase drift noise inside the mode-sorter interferometer. A phase modulator (PM$_c$) placed in one of the arms is used as the actuator for this control system.}
 \label{Fig3}
\end{figure*}

We now describe in detail the experimental setup for the two-dimensional version of our scheme (Fig. \ref{Fig3}). We choose to create the OAM$_{\pm1}$ modes through the well-known technique of generating a computer-generated hologram on a spatial light modulator (SLM) to mimic the desired optical element capable of converting a Gaussian laser beam into a helical mode \cite{Arrizon:07}. Although this is done in free-space for convenience, our mode-sorting system is completely in-fiber, and it would work equally well if the OAMs were generated with an in-fiber method \cite{Alarcon:23}. A continuous wave (CW) laser operating at $1546.92$ nm is spatially filtered through a single-mode fiber (SMF) and collimated with a 10x objective lens to reach a waist of $\omega_{p}\approx1250$ $\mu m$. A polarizing beam splitter (PBS), is used to set the laser's polarization state to horizontal for the optimal operation of the SLM. A manual polarization controller placed before the objective is used to optimize the optical power following the objective. Note that $\omega_{p}$ is sufficient to illuminate all the $\ell$-forked holograms, which consists of a helical phase profile superposed with a linear phase ramp to spatially isolate the encoded field from the Gaussian mode, resulting in a diffraction grating that produces the OAM$_{\ell}$ mode in the first diffraction order \cite{BEKSHAEV:10}. We use a Holoeye Pluto-Telco-013 phase-only SLM with a resolution of $1920$ x $1080$ pixels and $8\mu m$ pixel pitch to implement the holograms. Each pixel acts as a programmable phase shifter between $0$ and $2\pi$, which is controlled by the grayscale value of the corresponding pixels of the hologram.\\

The SLM reflects the beam onto a $4$f system composed of two lenses, L$_1$ and L$_2$, each with a focal length of $150$ mm. The Fourier transform of the reflected field is formed in the focal plane of L$_1$, where there is a pinhole. This spatial filter aims to pick out the diffracted first order, corresponding to the Fourier spectrum of the encoded field. The second lens (L$_2$) performs an inverse Fourier transform, resulting in the OAM$_{\pm1}$ modes. A beam splitter (BS) is placed at the end of the 4f system in order to split the beam and allow us to image the generated mode during the experiment. The transmitted mode is imaged by an InGaAs CCD camera placed in the image plane of L$_2$ (see Fig. \ref{Fig3} and inset). Both the amplitude and phase profiles of the OAM modes are imaged on the camera. For the phase profiles, we create a reference mode by splitting the laser's beam with a 50:50 fiber coupler before the input 10x objective (not shown in the figure for the sake of simplicity). This reference beam is connected to the fiber launcher named "Control input" in Fig. \ref{Fig3}, which consists of another 10x objective, where it is superposed with the OAM beam on the beamsplitter. Therefore, the phase profiles are imaged as the interference pattern between these two beams, and shown in the inset of Fig. 3. This control laser input is otherwise used as a launch device to a second diode laser as part of an active phase stabilization system for the mode sorter interferometer (further explained below).\\ 

On the reflected output of the BS there is an $20$x objective used to couple the OAM$_{\pm 1}$ mode into a FMF. The FMF used in this experiment is a commercially available graded index telecom fiber (OFS 80730) that can support three modes: the fundamental Gaussian LP$_{01}$ and the higher order LP$_{11a}$, and LP$_{11b}$ modes. The linearly polarized modes are particular modes that propagate in a weakly guided medium. Therefore, when the $20$x objective couples the OAM$_{\pm1}$ mode into the FMF, they are decomposed into LP modes propagating in the FMF. A manual fiber polarization controller, where the FMF is wound, is used to correctly adjust the modal demultiplexing in the photonic lantern. This fiber carrying the incoming OAM$_{\pm 1}$ modes is then connected to a commercial PL (Phoenix Photonics 3PLS-GI-15) consisting of one FMF at its input and three single-mode fibers at its output, corresponding to each of the supported LP modes. The internal structure of the PL is made of an adiabatic region that provides a low transition from the input FMF to the output of the three SMFs in such a way that each LP mode is mapped to the appropriate output port \cite{Leon-Saval:10, Birks:15}. This mapping is produced by a matching process between the effective indices of the tapered region and the incoming spatial modes \cite{Zeng:18, Li:19, Noordegraaf:09}. Since the incoming OAM mode only decomposes into the LP$_{11a}$ and LP$_{11b}$ modes, we will use only the two outputs of the PL corresponding to these modes leaving out the optical port that decomposes the LP$_{01}$ mode. After the OAM$_{\pm1}$ modes are decomposed, the following mapping is done: LP$_{11a}$→ $E_1$;  LP$_{11b}$→  $E_2$, where $E_1$ and $E_2$ represent both electrical fields propagating through the two different paths after PL.\\

A lithium niobate pigtailed telecom phase modulator (PM)(Thorlabs LN65S-FC) is placed on the upper arm. The PM is driven by an electrical signal from a function generator, so any arbitrary relative phase $\phi$ can be added to the upper arm between 0 and $2\pi$. A variable optical attenuator (not shown for simplicity) is located in the lower arm to equalize the optical power from both arms of the interferometer before recombining on the $50:50$ fiber coupler (FC). We also have manual polarization controllers before the FC to ensure path indistinguishability. D$_{+1}$ and D$_{-1}$ are amplified p-i-n photodiodes to measure the optical intensity at the two outputs, with the results recorded by a digital oscilloscope.\\

Due to the natural instability of the interferometric configuration, we inject a laser at a different wavelength ($1546.12$ nm) through the free port of the beam splitter before the InGaAs CCD camera (Fig. \ref{Fig3}). This laser beam co-propagates with the OAM modes through the sorter, and is used as a feedback signal for a phase stabilization system. This reference beam is split from the OAM information before D$_{+1}$ with a dense wavelength division multiplexer (DWDM), and is detected by a third amplified p-i-n photodiode D$_r$, whose electrical output is read by a microcontroller running a perturb-and-observe algorithm \cite{AlTaha:19} to stabilise the environmentally-induced phase drift in the mode sorter interferometer. The phase controller employs a second phase modulator (PM$_c$) placed in the opposing arm to compensate the phase drift. The control system operates continuously, but is briefly switched off when there is a need to operate the sorter in routing mode.  The optical loss from the PL input to the detectors is 8.3 dB, which are mainly given by the PM and the internal losses of the PL ($\sim$ 5 dB). These losses can be greatly reduced by new lantern designs that have much lower losses (up to 0.2 dB) \cite{Guo:23}, showing a path to having a mode sorter with very low losses.

\section{Results}\label{sec:Results}

\begin{figure}[!ht]
\centering
\includegraphics[width=0.5\textwidth]{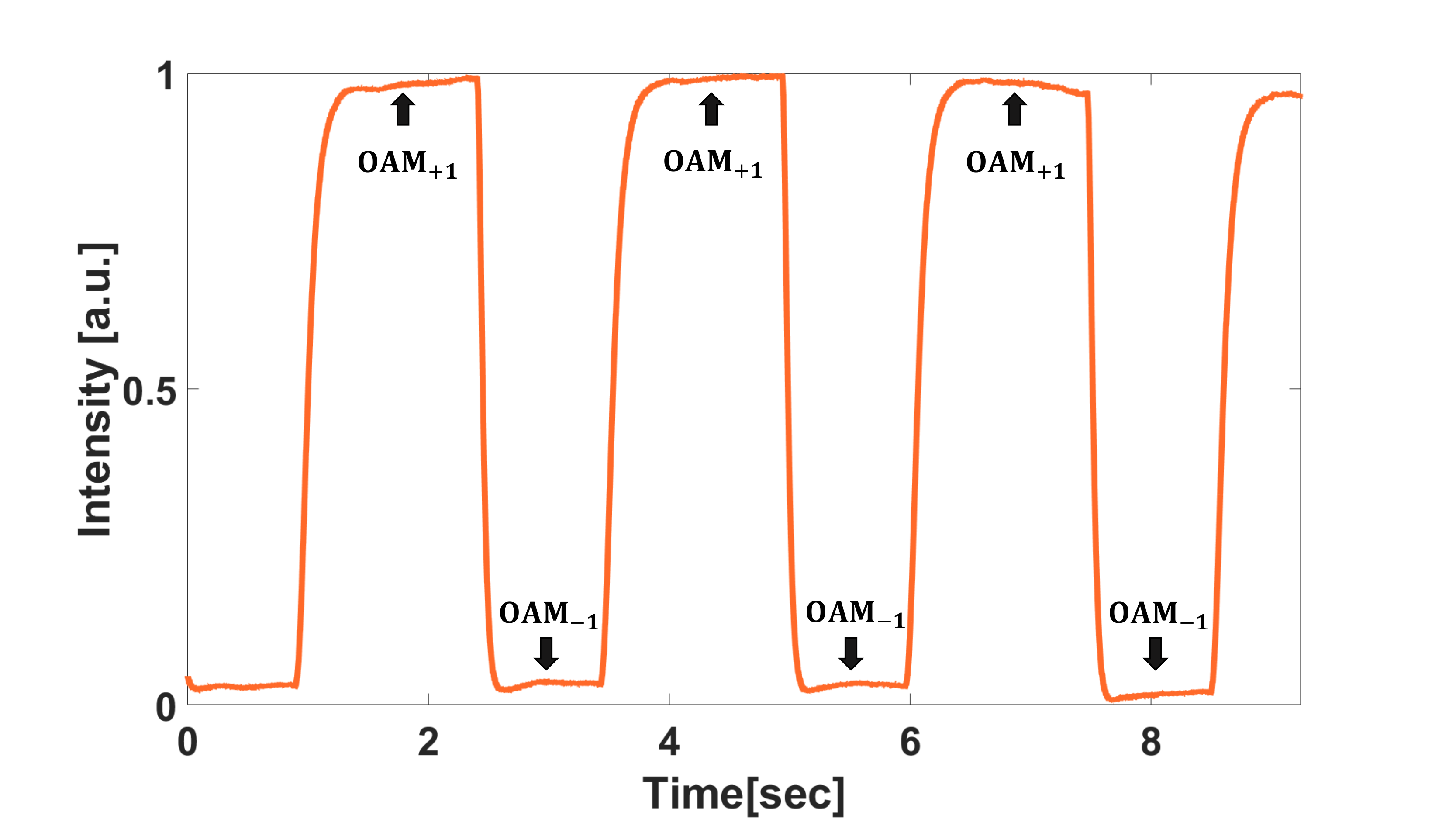}
\caption{OAM mode sorting. Here we show how our setup can sort the OAM modes when the source is constantly changing between OAM$_{+1}$ and OAM$_{-1}$ with a period of 1 s. A relative phase of $\pi/2$ is being applied inside the interferometer by PM, such that OAM$_{+1}$ is sent to D$_{+1}$ and OAM$_{-1}$ is sent to D$_{-1}$. The intensity curve above as a function of time is measured at D$_{+1}$. The OAM mode labels indicate when each particular OAM mode was being sent.}
\label{Fig4}
\end{figure}

The first goal is to demonstrate the successful sorting between the OAM$_{\pm 1}$ modes with our scheme. Given that the mode sorter Mach-Zehnder interferometer (MZI) is actively stabilized by the controller, the power detected in D$_{+1}$ and D$_{-1}$ will depend on the additional relative phase $\phi$ applied by the PM and also of the topological charge $\ell=\pm 1$ of the incoming OAM beam. It is well known that in an MZI the optical intensity registered in $D_{+1}$ ($D_{-1}$) is proportional to ${|E_1+e^{i\phi}E_2|}^2$ (${|E_1-e^{i\phi}E_2|}^2$). From Eq. \ref{eq1}, we notice that the incoming  OAM$_{\pm 1}$ also carries a relative phase between both LP modes of $\pm \pi/2$. By adjusting the controller to maintain a relative phase between both arms of $\pi/2$, then if the incoming beam is an OAM$_{+1}$ (OAM$_{-1}$) then we obtain maximum optical intensity detected at $D_{+1}$ ($D_{-1}$). We then continuously switch between OAM$_{+ 1}$ and OAM$_{- 1}$ every second by changing the corresponding forked diffraction gratings in the SLM. Figure \ref{Fig4} shows the electrical signal corresponding to the optical intensity measured at D$_{+1}$,  when fixing $\phi =\pi/2$ in the interferometer, with the periodic switching of the OAM modes. From this result we show the deterministic identification of the direction of rotation of the OAM mode, thus its topological charge.

\begin{figure}[htp]
\centering
\includegraphics[width=0.45\textwidth]{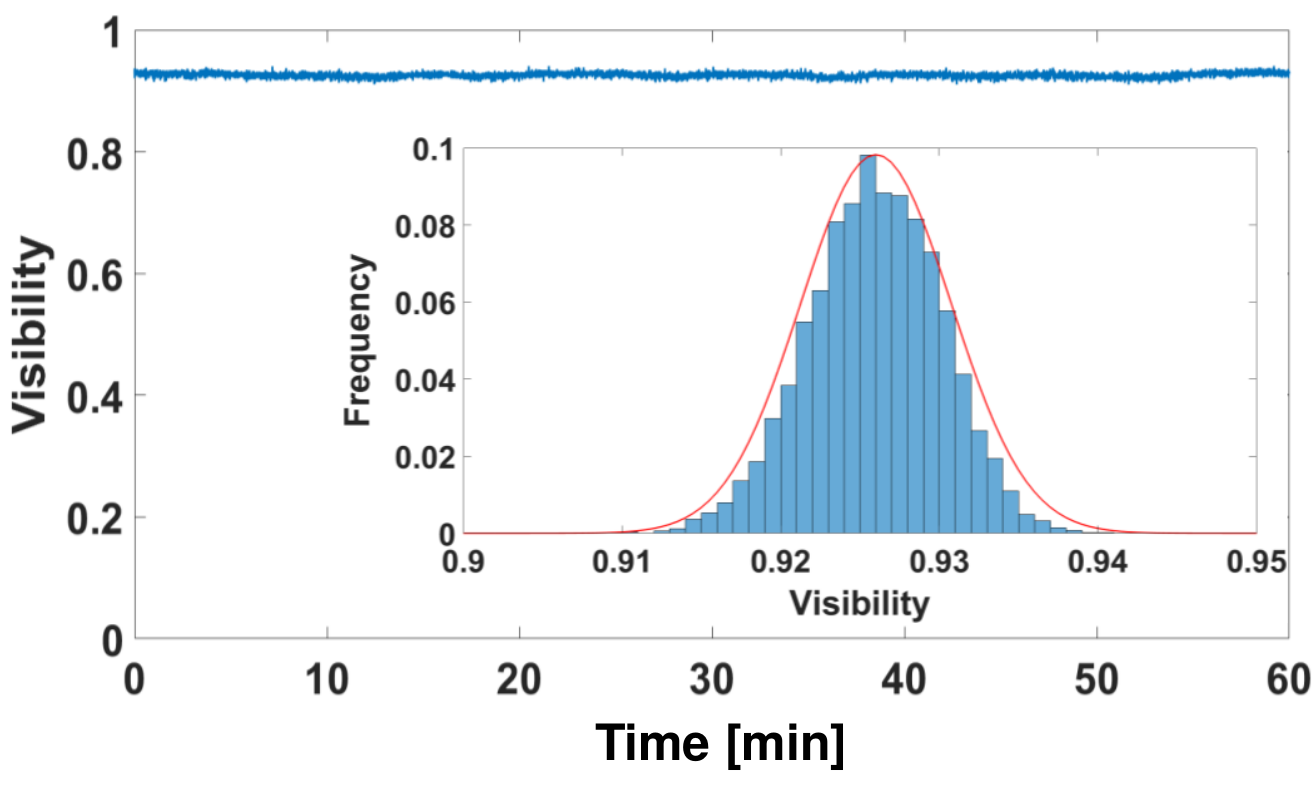}
\caption{Long-term performance of the OAM mode sorter. We plot the visibilities during a period of one hour calculated from the constant switching between the OAM$_{+1}$ and OAM$_{-1}$ modes from the source, through the SLM. The inset shows a histogram comes from these results, showing that the average visibility is 92.59$\pm 1.87$\%.}
\label{Fig5}
\end{figure}

To demonstrate the stability of the sorting process over longer time periods, we have repeated this process continuously for one hour. We have calculated the optical visibility using the consecutive optical intensities between the maxima and minima at the output of D$_{+1}$ as the input OAM mode is continuously switched. For any two consecutive maxima and minima, we calculate the visibility $V$ as $V=(I_{+}-I_{-})/(I_{+}+I_{-})$, where $I_{+}$ ($I_{-}$) is the detected optical intensity when the OAM$_{+1}$ (OAM$_{-1}$) mode is sent. We plot the visibilities calculated for every mode transition as a function of time in Fig. \ref{Fig5}, while also displaying in the inset, the histogram of all the measured visibilities giving an average of $92.59 \pm 1.87 \%$.\\

\begin{figure*}[!ht]
\centering
\includegraphics[width=1\textwidth]{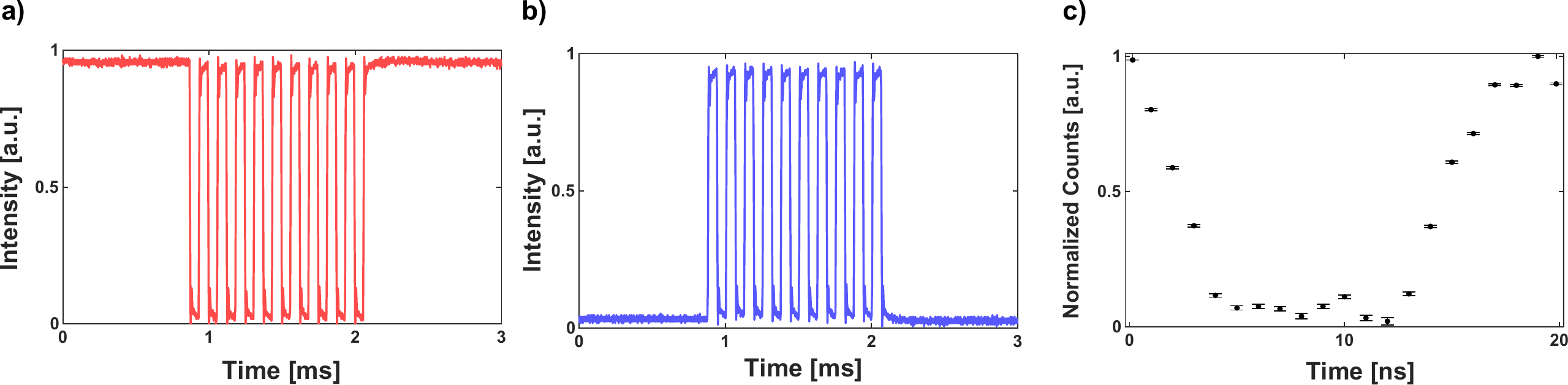}
\caption{OAM mode routing with the source transmitting the OAM$_{+ 1}$ state. a) Optical intensity at the D$_{+1}$ output when the internal PM in the mode sorter changes $\phi$ from $\pi/2$ to $-\pi/2$ ten times consecutively in a period of 1.1 ms, generated from a field programmable gate array (FPGA) electronic module. During the switching time, the phase stabilization algorithm is suspended, and resumed after the last switching pulse is applied. In this way, it is possible to route the OAM$_{+ 1}$  mode to any detector arbitrarily. b) The same measurement, but observed in detector D$_{-1}$, where the complementary nature of the interferometer can be observed, as expected. The routing speed in measurements (a) and (b) are limited by our p-i-n photodetectors. In (c) we demonstrate the speed response of our mode sorter by changing the p-i-n photodiode at D$_{+1}$ for an InGaAs single-photon detector, which has a timing resolution of 200 ps. For this measurement we have attenuated the output power of the source to a few photons per detector gate window of 2.5 ns. We measure a rise/fall time of approximately 7 ns, limited by our employed driving electronics.}
\label{Fig6}
\end{figure*}

Our system can also operate as an active router for OAM modes by applying fast relative phase changes through the PM. For this demonstration we fix the OAM$_{+ 1}$ in the source while electrically driving a train of ten pulses, each being 100 $\mu$s long, to the PM. During the period the pulse train is being sent the phase stabilization algorithm is suspended. The amplitude and offset of these driving pulses was previously calibrated in order to have a relative phase applied between the paths equivalent to $\phi=\pi/2$ and $\phi=-\pi/$2 respectively. We once again observe the output of the system by recording the optical intensity present at D$_{+1}$. By applying the sequential pulses to the PM, the OAM$_{+ 1}$ mode can be routed to either D$_{+1}$ or D$_{-1}$ depending on the value of the driving relative phase, with these results shown in Fig. \ref{Fig6}(a), where we can clearly see the sequential routing of the OAM$_{+ 1}$ mode between the two outputs. The system works equally well if we now observe the D$_{-1}$ output as shown in Fig.\ref{Fig6}(b), with the routing operation complementary as expected. The average visibility for both cases is $92.75\pm 1.27\%$, with the main limitation to the optical visibility originating from the modal crosstalk in the photonic lantern.\\

The ultra-fast reconfigurability is a major advantage of our configuration stemming from the use of telecom electro-optical fiber pig-tailed phase modulators. For this demonstration, we replace the p-i-n photodiode, which has a limited bandwidth of 10 MHz, with an InGaAs single-photon counting module (IdQuantique id210), which has a timing resolution of 200 ps. The detector is running with 10\% detection efficiency and a 2.5 ns wide gate window. We then modulate the internal PM in the mode sorter with a series of 20 ns wide transitions between $\pi/2$ and $-\pi/2$ and synchronize the detector's gate window with these transitions. We then vary the electrical delay to the trigger signal fed to the InGaAs detector in steps of 1 ns, allowing us to scan the optical phase pulse which switches between the OAM modes, and the result is shown in Fig. \ref{Fig6}(c), where we show a rise/fall time of $7$ ns, which is limited by our home-made electrical driver amplifying the electrical signal feeding the phase modulator. This value can be further improved with appropriate electronics, as the response bandwidth of the phase modulator is within the 10 GHz range. For this measurement the OAM laser beam is attenuated such that only a few photons per detection window are present and thus, the detector is not saturated.

\section{Conclusions}\label{sec:Conclusion}

Appropriate tools for manipulating OAM states are highly sought in many areas of photonics, opening up new possibilities based on this degree-of-freedom of light. Sorting of OAM modes has been one particular area of considerable intense research in the last years, using many different techniques. Common to most of these previous methods is the use of bulk optical elements to a greater or lesser degree. The downside of these schemes is the difficulty of integration with optical fiber systems, due to the additional losses and complexity needed to couple the light to and from a fiber. This is specially acute in cases where long-distance transmission is desired, such as in optical and quantum communications. In the few cases where optical-fiber-based techniques were used, no reconfigurability or high speed responses were possible, severely limiting those to applications where dynamic operation is essential, such as in quantum information processing. Here we solve this issue by combining photonic lanterns with interferometric analysis of the components of an OAM mode. Our sorter is designed to receive an OAM mode coming from a few-mode fiber, and decomposes it into its linearly polarized mode components with a photonic lantern. Then by recombining these modes interferometrically onto a beamsplitter, we are able to sort the incoming modes. Our scheme is generalizable to higher dimensionalities by employing few-mode fibers, photonic lanterns and multi-port beamsplitters that support more transverse spatial modes. For instance,  MBSs with 7 ports \cite{Carine:20} or alternatively arrays of 2 port beamsplitters on a photonics integrated circuit can function as a general $d$-port beamsplitter \cite{Clements:16} can be used as well as lanterns with 15\cite{Fontaine:15}, 19 \cite{Noordegraaf:12} and 61 \cite{Noordegraaf:10} ports have been demonstrated, showing the potential of our scheme. Furthermore very recent results have demonstrated photonic lanterns with very low loss ($<$ 0.2 dB) \cite{Guo:23}, further showing the feasibility of our proposal for practical applications.\\

Finally, due to the interferometric nature of our configuration, it can also be used in an active/reconfigurable mode, where the input mode can be routed to a different output on demand, or a unitary transformation can be applied by the phase modulator, allowing the measurement in different mutually unbiased bases in quantum information \cite{Carine:20}. In this active mode, we have demonstrated, to be the best of our knowledge, the fastest response in an OAM router, with the potential to reach sub-ns response times due to the use of fast telecom phase modulators. Our results show how the use of space-division multiplexing technology (i.e. few-mode fibers and photonic lanterns) can be used to implement a fast and reconfigurable OAM mode sorter that is fully compatible with optical fiber networks, further paving the way for this popular degree-of-freedom to be used as a reliable transport medium for information in classical and quantum networks.

\subsection*{Acknowledgements}
The authors acknowledge support from CENIIT Link\"{o}ping University, the Swedish Research Council (VR 2017-04470), the Knut and Alice Wallenberg Foundation through the Wallenberg Center for Quantum Technology (WACQT) and by the QuantERA grant SECRET (VR 2019-00392). Also, this work was supported by the Fondo Nacional de Desarrollo Científico y Tecnológico (FONDECYT) (Grant Nos., 1230796, 11201348 and 1200859) -- Millennium Science Initiative Program -- ICN17$_-$012.

%%%%%%%%%%%%%%%%%%%%%%% References %%%%%%%%%%%%%%%%%%%%%%%%%

\bibliography{main}

\end{document}